\documentstyle[prl,aps,epsf,multicol]{revtex}

\begin{document}

%
\newcommand{\fig}[2]{\epsfxsize=#1\epsfbox{#2}}
%
%
%


 \newcommand{\passage}{
         \end{multicols}\widetext%
                \vspace{-.5cm}\noindent\rule{8.8cm}{.1mm}\rule{.1mm}{.4cm}}
 \newcommand{\retour}{
         \vspace{-.5cm}\noindent\rule{9.1cm}{0mm}\rule{.1mm}{.4cm}\rule[.4cm]{8.8cm}{.1mm}%
         \begin{multicols}{2} }
 \newcommand{\unecol}{\end{multicols}}
 \newcommand{\deuxcol}{\begin{multicols}{2}}
%

\tolerance 2000

\author{Baruch Horovitz{$^1$} and Pierre Le Doussal{$^2$} }
\address{{$^1$} Department of Physics, Ben Gurion University, Beer Sheva
84105 Israel}
\address{{$^2$}CNRS-Laboratoire de Physique Th{\'e}orique de
l'Ecole Normale Sup{\'e}rieure, 24 rue Lhomond,75231 Cedex 05,
Paris France}
\title{Freezing transitions and the density of states of 2D random Dirac Hamiltonians}
\maketitle

\begin{abstract}
Using an exact mapping to disordered Coulomb gases, we introduce a
novel method to study two dimensional Dirac fermions with quenched
disorder in two dimensions which allows to treat non perturbative
freezing phenomena. For purely random gauge disorder it is known
that the exact zero energy eigenstate exhibits a freezing-like
transition at a threshold value of disorder
$\sigma=\sigma_{th}=2$. Here we compute the dynamical exponent $z$
which characterizes the critical behaviour of the density of
states around zero energy, and find that it also exhibits a phase
transition. Specifically, we find that $\rho(E=0 + i \epsilon)
\sim \epsilon^{2/z-1}$ (and $\rho(E) \sim E^{2/z-1}$) with $z=1 +
\sigma$ for $\sigma < 2$ and $z=\sqrt{8 \sigma} - 1$ for $\sigma >
2$. For a finite system size $L<\epsilon^{-1/z}$ we find large
sample to sample fluctuations with a typical $\rho_{\epsilon}(0)
\sim L^{z-2}$. Adding a scalar random potential of small variance
$\delta$, as in the corresponding quantum Hall system, yields a
finite noncritical $\rho(0) \sim \delta^{\alpha}$ whose scaling
exponent $\alpha$ exhibits two transitions, one at $\sigma_{th}/4$
and the other at $\sigma_{th}$. These transitions are shown to be
related to the one of a directed polymer on a Cayley tree with
random signs (or complex) Boltzmann weights. Some observations are
made for the strong disorder regime relevant to describe transport
in the quantum Hall system.
\end{abstract}

\deuxcol


\section{introduction}

The critical behaviour of the plateau transitions in the
integer quantum Hall (QH) effect remains an appealing
theoretical challenge. Despite numerous attempts, a calculable
theory remains elusive. An equivalent version of the quantum Hall system
which is believed to capture the relevant physics
corresponds to two dimensional Dirac fermions in presence of {\it } both a random vector and
a random scalar potential \cite{ludwig}. Conventional perturbative methods have failed and it is believed that
the problem is described by some non perturbative strong coupling regime \cite{ludwig,bernard}.
Recent works using conformal field theory \cite{zirnbauer,leclair,leclair2,mudry,gurarie}
or non linear sigma models aim at reaching this regime \cite{nlsigma}.

One possible route of attack is to use the boson representation
 \cite{ludwig,lee,marston}
based on the network model \cite{chalker}. Indeed, the model can
be mapped exactly, via bosonization, onto a random sine Gordon
model or equivalently a Coulomb gas (CG) with a specific type of
disorder. Although the calculation of the density of states, via
the retarded Green's function, corresponds to considering a single
CG layer, the full treatment of the quantum Hall transition (both
advanced and retarded Green's function) requires to study two
coupled Coulomb gas layers and remains highly non trivial in these
new variables. On the other hand, there has been recent progress
in understanding disordered CG, mainly in the context of random
gauge XY models
\cite{nattermann,kor_nat,tang_xy_lowtemp,scheidl,carpentierxy,carpentierpart,h}
and in particular the freezing transitions which occur in these
systems. New methods, such as fugacity distribution
renormalization group (RG) \cite{carpentierxy,carpentierpart,giam}
as well as variational methods \cite{h}, have been developed which
seem to be able to capture some of the non perturbative features
of the strong disorder regimes. It is thus of interest to search
what can be learned from these methods and to understand,
irrespective of formal technicalities, whether the (glass
transition) physics that they describe will be part of the QH
strong disorder physics.

In this paper we mainly focus on the detailed understanding of the single layer problem in
the Coulomb gas formulation with the practical aim of computing the density of states.
We also extend our method to the full QH problem, proposing a new approach
on this venerable problem.

We start by further restricting to the purely random vector
potential disorder model, the scalar random potential will be
added later on. This simpler model has been intensively studied
\cite{ludwig,gurarie,mudry,chamon,tsvelik_prl} and is believed to
be critical, with a line a fixed points, and a continuously
varying dynamical exponent $z(\sigma)$ as a function of random
vector potential disorder strength $\sigma$. Some precise results
exist for an exactly known zero energy eigenstate which has the
form $\psi(x) = e^{U(x)/2} \psi_0$ where $U(x)/2$ is the primitive
of the vector potential. It was found \cite{castillo} that
averaged moments scale with system size $L$ as $\sum_x
|\psi(x)|^{2q} \sim L^{-\tau (q)}$, such that above a threshold
value $\sigma=\sigma_{th}$ of disorder $\tau (q)=0$ for
sufficiently large $q$ indicating some kind of localized
behaviour. Further studies \cite{carpentierpart} confirmed the
existence of a transition at $\sigma=\sigma_{th}$ in the (Gibbs
like) probability measure $|\psi(x)|^2=e^{U(x)}$ (equivalently a
freezing, i.e. a glass transition), as well as its relations, via
RG, to the directed polymer on a Cayley tree
\cite{derridaspohn,tang_xy_lowtemp,chamon} and found a non trivial
structure of the strong disorder phase with ''quasi localized''
behaviour. Interesting relations to the Liouville theory,
conjectured in \cite{tsvelik_prl} were reexamined and it was found
that the freezing transition can be directly demonstrated from
renormalization in the Liouville model \cite{carpentierpart}.

The known results about the exact $E=0$ eigenstate \cite{castillo}
do not however tell
anything directly about the density of states. In particular the dynamical
exponent has not yet been calculated in the strong disorder regime,
and one would guess that it should exhibit some
kind of change at the transition $\sigma=\sigma_{th}$. A freezing in the
dynamical exponent was indeed demonstrated recently \cite{freezing} in a closely related model,
i.e. the classical Arrhenius diffusion in the potential $U(x)$, in both one and two dimension
at the same value $\sigma=\sigma_{th}$ than the $E=0$ eigenstate transition.
In one dimension the square of
the Dirac Hamiltonian is well known to be identical to the Fokker Planck operator
and there the two problems are thus equivalent \cite{old}. Thus in one dimension, if one considers a
log correlated $U(x)$, both problems have {\it identical} dynamical exponents and freezing transitions
given in \cite{freezing}. In two dimensions, as discussed below, the two differ
by a an additional imaginary random drift term, but still they both have a line of fixed
points and it is reasonable that they would both undergo freezing transitions,
as we find here.

In this paper we start by defining the models (Section II) and by
showing that the density of states (DOS) of the Dirac Hamiltonian
can be expressed as an observable in a boson formulation. For
convenience we study the DOS $\rho_{\epsilon}(E)$ at energy $E=0$
adding a small but finite imaginary $i \epsilon$ term for the
retarded propagator, thus in effect computing a smoothed DOS, and
carefully study the limit $\epsilon \to 0^+$ (Section III). At
$E=0$ the model becomes very similar, in the boson formulation, to
the random gauge XY model in the phase where vortices are
relevant. The parameter $\epsilon$ plays the role of a bare vortex
fugacity and the local DOS $\rho_{\epsilon}(0,{\bf r})$
corresponds to the renormalized vortex fugacity $z_{\pm}({\bf r})$
(or the local density) which becomes broadly distributed when
$\epsilon \to 0$. We show that the order of limits
$\epsilon\rightarrow 0$ and system size $L\rightarrow \infty$ is
significant. For $L\rightarrow \infty$, such that the typical
level spacing $\Delta E<\epsilon$, we use a variational scheme and
show that $\rho_{\epsilon}(0) \sim \epsilon^{2/z-1}$ with $z$
exhibiting a transition at a critical value of disorder. This is
equivalent to a phase transition in  $\rho(E)\sim E^{2/z-1}$. For
$\Delta E> \epsilon$ we find that $\rho_{\epsilon}(0)$ becomes
analogous to to the partition function of a directed polymers on a
Cayley tree, and also exhibits the freezing transition. It is
however a strongly fluctuating quantity in that limit and is
interpreted as a typical value, rather than a disorder average.
Further analogies with freezing of dynamical exponents in
Arrhenius dynamics is presented. Finally, in Section IV we include
a scalar random potential as in the full Quantum Hall system. We
find that the DOS is non critical in $E$, however its $\sigma$
dependence is critical. We also develop a variational scheme for
studying the transport and localization exponents.

\section{single layer model, definitions, and exact mappings}

Our aim is first to study the density of states of the random Dirac Hamiltonian
in two space dimensions:
\begin{eqnarray}
H_D =  \hbar v_F {\mbox{\boldmath
$\sigma$}} \cdot [-i{\mbox{\boldmath $\nabla$}} -{\bf A}({\bf
r})] + W({\bf r})
\label{hdirac}
\end{eqnarray}
where ${\bf r}=(x,y)$ is the 2D space, $\sigma =(\sigma _x, \sigma_y)$
are Pauli matrices, $W({\bf r})$ is a random scalar potential and ${\bf A}({\bf r})$ is a random vector
potential (in units of $e/\hbar$), both gaussian with short range correlations (in
the following we set $\hbar v_F=1$). ${\bf A}$ can be chosen
purely transverse ${\bf A}_x=\partial_y V$, ${\bf A}_y=- \partial_x V$
and its potential has logarithmic correlations $\overline{(V({\bf r}) - V({\bf r}'))^2}
\sim \sigma \ln|{\bf r} - {\bf r}'|$, which defines $\sigma$. Two exact zero energy
(unnormalized) eigenstates are then $\psi=(e^{V},0)$ and $\psi=(0,e^{-V})$.

Since we are also interested in the local density of states in a given
sample and that this is a
fluctuating quantity it is convenient to
define the smoothed local density of states $\rho_{\epsilon}(E,{\bf r})$
as:
\begin{eqnarray}
&& \rho_{\epsilon}(E,{\bf r})=\frac{1}{\pi} \Im
< {\bf r} | \frac{1}{E - H_D - i\epsilon} | {\bf r} > \nonumber \\
&& = \frac{1}{\pi} \sum_n
 \frac{\epsilon}{(E-E_n)^2 + \epsilon^2} |\psi_n({\bf r} )|^2 \label{rho}
\end{eqnarray}
where $E_n$ are energy eigenvalues of $H_D$ and $\psi_n$ the eigenstates. For $\epsilon \to 0^+$ one recovers
the standard local DOS, and for finite $\epsilon$ each level is broaden by a Lorentzian.
The standard DOS is then defined as the spatial average, for a system of linear
size $L$:
\begin{eqnarray}
&& \rho_{\epsilon}(E) = \frac{1}{L^2} \int d^2 {\bf r} \rho_{\epsilon}(E,{\bf r})
\end{eqnarray}
Although this usually becomes a smooth function of $E$ for $L \to  \infty$ in any given sample,
at finite size and for small $\epsilon$ it is a series of peaks whose locations usually fluctuate strongly
from sample to sample. Clearly these fluctuations are smoothed when $\epsilon$ becomes
of the order or larger than the typical level spacing $\Delta E$.
Naively, if the lowest energy states scale as $L^{-z}$ then
dimensional argument gives for the space averaged DOS
 $\rho_{\epsilon}(0)\sim L^{z-2}$ for $\epsilon<L^{-z}$ or
$\rho_{\epsilon}(0)\sim \epsilon^{2/z-1}$ for $\epsilon>L^{-z}$.

The local DOS can be expressed from the free fermion
action with spinors $\overline{\psi}({\bf r})$, $\psi({\bf r})$,
projected into a subspace of energy $E$ which defines the
Dirac problem in $2+1$ dimensions.
\begin{eqnarray}
&& \rho_{\epsilon}(E,{\bf r}) = \frac{1}{\pi} \Im < \overline{\psi}(r) \psi(r) >_{{\cal S}_D} \\
&& {\cal S}_D = \int d^2r \overline{\psi}({\bf r})\{{\mbox{\boldmath
$\sigma$}} \cdot [-i{\mbox{\boldmath $\nabla$}} -{\bf A}({\bf
r})]  \nonumber \\
&&+W({\bf r})-E+i\epsilon \}\psi({\bf r}) \label{SD}
\end{eqnarray}
An additional Dirac mass term $\Delta_d \sigma_z$ in Eq.
(\ref{SD}) controls the distance from criticality and is set here
to zero.

The problem can be mapped onto a sine Gordon model.
Considering $y$ as an imaginary time variable this action
can be written as a 1+1 dimensional fermion problem. Further
bosonization \cite{ludwig} yields the action

\begin{eqnarray}
{\cal S}_B = &&\int d^2r\{ \frac{1}{ 8 \pi K} [{\mbox{\boldmath
$\nabla$}} \theta({\bf r})]^2 + \frac{i}{2 \pi} [A_y({\bf r})\partial_x -A_x({\bf
r})
\partial _{y}] \theta ({\bf r}) \nonumber\\
&&- \frac{i}{\pi \alpha} (W({\bf r})-E+i\epsilon) \cos \theta({\bf r}) \}
\label{SB}
\end{eqnarray}
where $\alpha$ (which denotes $\hbar v_F \alpha$) is the momentum cutoff, and $K=1$; $K\neq 1$ may
be generated by RG or correspond to 1D type interactions.
Allowing for $K\neq 1$ is mainly instructive as it allows to interpolate towards the
random gauge XY model, and we call that situation the generalized Dirac
model. Eq. (\ref{SB}) can also be derived from the network
model of the QHE \cite{lee,marston,chalker} where x is discretized,
$W(x,\tau)$ is then the long range component of the random
potential while $A_y(x,\tau)=(-)^x W(x,\tau)$ is the short
wavelength component; both terms couple to slowly varying fields,
hence can be considered as independent random variables, though
with equal averages.

In the sine Gordon formulation, the local (smoothed) DOS is
given exactly by an average of the operator $\cos(\theta)$
as follows:
\begin{equation}
\rho_{\epsilon}(E,{\bf r})= - \frac{1}{\pi
L^2}\frac{\delta}{\delta W({\bf r})}\Im \ln Z =\frac{1}{\pi ^2\alpha}
 \Re \langle cos\theta({\bf r})\rangle
 \end{equation}
where $Z=\int D \overline{\psi} D \psi e^{-{\cal S}_D} $
and $\langle cos\theta\rangle$ is an average over $\theta$
with the action Eq. (\ref{SB}).

We note that all the above mappings are exact. They are even exact for a
finite size sample with some specified boundary conditions for the path integrals.
We will not need here to detail the correspondences in boundary conditions, but it
can be done in principle. Note that since the action is complex, the
$\langle cos\theta({\bf r})\rangle$ can be arbitrarily large (when the
denominator vanishes) which is the case for $\epsilon \to 0^+$ as $E$
crosses an eigenvalue $E_n$.

\section{Random vector potential model}

We now set $W({\bf r})=0$ and study the model with only
a random vector potential. To determine the dynamical exponent $z$
we will study the smoothed DOS at zero energy $E=0$. Below
we will distinguish two limits and study them separately.
First in the large size limit ($\epsilon > L^{-z}$), if
we assume, as is customary, that there is a well defined density of states
$\rho(E) \sim E^{2/z-1}$, when $L \to + \infty$ one has:
\begin{eqnarray}
\rho_{\epsilon}(0) = \frac{1}{\pi} \int dE \rho(E) \frac{\epsilon}{\epsilon^2 + E^2}
\sim \epsilon^{2/z-1}
\label{integral}
\end{eqnarray}
for fixed small $\epsilon$ and $z>1$. Thus we can obtain $z$ unambiguously from $\rho_{\epsilon}(0)$.
This observable should be self averaging in that limit since the DOS at zero energy receives
contributions from many energy levels in a window of size $\epsilon$ around $E=0$,
and this is what we find below.

There is another interesting limit, also studied below, when
$\epsilon < L^{-z}$ is small (respectively finite size). Then
there are fewer energy levels and $\rho_\epsilon(0)$ becomes a
strongly fluctuating quantity, as discussed below, which gives information
about the statistics of the lowest energy levels $E_0$ near $E=0$,
and thus also about the typical energy level spacing, found to scale
also as $E_0 \sim L^{-z}$ with the same exponent $z$.

\subsection{Large size limit}

A look at Eq. (\ref{SB}) setting $E=0$ (with $W({\bf r})=0$) shows
that the model is {\it identical} to the sine Gordon (or
equivalently Coulomb gas) formulation \cite{carpentierxy,h} of the
random gauge XY model. Clearly $\epsilon$ plays the role of the
bare vortex fugacity, but as shown in \cite{carpentierxy} the
random vector potential generates, upon coarse graining, an
additional local random potential resulting in a random fugacity
$z_\pm({\bf r}) = \epsilon e^{ \pm U({\bf r})}$ for $\pm 1$
charges. Some of the physics of the random gauge XY model will
thus be relevant here. In particular the local DOS
$\rho_{\epsilon}(0,{\bf r})$ is analougous to the coarse grained
$z_\pm({\bf r})$ and thus become broadly distributed as $\epsilon
\to 0$, as discussed below.

It is convenient to perform a replica average on ${\bf A}$ in Eq. (\ref{SB}). This yields
a Hamiltonian for the replicated field $\theta_a({\bf r})$ with
replica indices $a,b=1,...,m$
\begin{eqnarray}
&& {\cal H}=\int d^2r\{(1/8\pi)\sum_{ab}(K^{-1}\delta_{ab}+\sigma)
{\mbox{\boldmath $\nabla$}} \theta _a \cdot {\mbox{\boldmath
$\nabla$}} \theta _b \\
&& -\sum_{{\bf n}}Y[{\bf n}]\exp (i{\bf n} \cdot
{\mbox{\boldmath $\theta$}})\} \label{H}\,.
\end{eqnarray}
where $\langle A_x^2(q) \rangle=\langle A_y^2(q)
\rangle=\pi\sigma$, ${\bf n}$ is a vector of length $m$ with
entries $0,\pm 1$ and $Y[{\bf n}]\sim \prod_a \epsilon ^{n_a^2}$.
The term $\epsilon \cos (\theta)$ in Eq. (\ref{SB}) corresponds to
$Y[{\bf n}]$ with $\sum_an_a^2=1$ while all other ${\bf n}$ are
generated by renormalization group (RG). The inclusion of all
these terms is essential for treating properly the strong disorder
situation \cite{scheidl,carpentierxy,h}, and obtaining the correct
scaling dimension of the $\epsilon \cos (\theta)$ operator, which
is what we need here. Since $\epsilon$ is finite, and we are
mostly interested in the region $K > 1/2$ where the vortices are
relevant, there will exist in finite density (separated by a scale
$\epsilon^{-z}$). Since we are interested in the end in the
behaviour as $\epsilon \to 0$ (dilute limit) we can use the RG
method developped in \cite{carpentierxy} and follow the full
distribution of fugacities or equivalently all the $Y[{\bf n}]$,
up to the length scale at which the vortices separation becomes of
order one, corresponding to $L \sim \sigma_c^{-1/2}$ below (see
Ref \cite{giam} for a similar RG study).

For simplicity we use instead a variational method, shown in our
previous studies \cite{h} to be good enough to describe the dilute
vortex system. In the limits of interest in Sections III and IV A
(small $E,\epsilon$) this variational method is easily seen (by
comparison to the above mentioned RG) to give the {\it exact
result } for scaling dimensions. Very much as in
\cite{carpentierxy} we expect the more precise RG treatment to
correct only amplitudes at weak disorder, powers of logarithmic
prefactors at strong disorder and be necessary mostly for detailed
descriptions very near the transitions, which we leave for future
publication. The results given below for the exponents $z$ and
$z'$ should thus be considered as exact.

The variational ${\cal H}_0$ has the form
\begin{eqnarray}
&& {\cal H}_0 =\int d^2r \{(1/8\pi)\sum_{ab}[(
K^{-1}\delta_{ab}+\sigma){\mbox{\boldmath $\nabla$}} \theta
_a \cdot {\mbox{\boldmath $\nabla$}} \theta _b \\
&& +(\sigma _c\delta
_{ab} +\sigma_0)\theta _a \theta _b\} \label{H0}
\end{eqnarray}
with $\sigma_c$, $\sigma_0$ variational mass parameters.
The propagators of the $\theta$ field are used to define
\begin{eqnarray}
&& \sum_{{\bf q}}\langle\theta_{a}({\bf q})\theta_{b}
(-{\bf q})\rangle_0 = -2u\delta_{ab}-A\nonumber\\
u&=&-(K/2)\ln (\Delta_c/4\pi K\sigma_c)\nonumber\\
A&=& \sigma K^2\ln(\Delta_c/4\pi K\sigma_c)+K\sigma_0/\sigma_c -
\sigma K^2 \,.
\end{eqnarray}
The interaction term is ($\epsilon/\pi \alpha\rightarrow\epsilon$
here)
\begin{eqnarray}
&& \langle \sum_{{\bf n}}Y[{\bf n}]\exp (i{\bf n} \cdot
{\mbox{\boldmath $\theta$}})\rangle \\
&&=
\langle\sum_{{\bf n}}
\exp[(u + \ln \epsilon) \sum_an_a^2+\omega\sum_an_a]\rangle_{\omega}
\label{av}
\end{eqnarray}
where the $\omega$ average reproduces the required form with
$A\sum_a n_a^2$,
\begin{equation}
\langle ...\rangle_{\omega} =\int ...\exp [-\omega^2/2A]
d\omega/(2\pi\sqrt{A})  \,.
\end{equation}
and has the physical interpretation of an average over random
local fugacities \cite{h,carpentierxy}. The sum in Eq. (\ref{av})
can be written as
 $\langle{\cal H}_{int}\rangle_0=\langle Z^m\rangle_{\omega}$
with
\begin{equation}
Z=1+\epsilon e^{u+\omega}+\epsilon e^{u-\omega}
\end{equation}

The variational free energy is then minimized,
$F_{var}=F_0 + \langle H-H_0 \rangle_0$, where $F_0$ is the
free energy of Eq. (\ref {H0}) and $\langle  \rangle_0$ is an average
with weights $\exp(-{\cal H}_0)$. This procedure yields \cite{h}
an equation for $\sigma_c$
\begin{equation}
\sigma_c = \int d\omega \,\frac{\epsilon e^{u+\omega}+\epsilon
e^{u-\omega}+4\epsilon^2e^{2u}}{(1+\epsilon e^{u+\omega}+\epsilon
e^{u-\omega})^2} \,e^{-\omega ^2/2A}\frac{d\omega}{\sqrt{2\pi A}}
\label{sigmac}
\end{equation}
where
\begin{eqnarray}
u&=&(K/2)\ln (4\pi K\sigma_c/\Delta_c)
  \nonumber\\
 A&=& -\sigma K^2ln(4\pi K\sigma_c/\Delta_c)-K^2\sigma
 +K\sigma_0/\sigma_c \label{uA}
\end{eqnarray}
 and $\Delta_c\gg \sigma_c$ is an integration cutoff. Eq.
 (\ref{sigmac}) can be solved by steepest descent  when the
 logarithms are large. A similar equation for $\sigma_0$ yields
 that $\sigma_0/\sigma_c$ in Eq. (\ref{uA}) is at most finite and
therefore can be neglected to determine exponents. The result is a
phase diagram shown in Fig. 1 with a massive phase $\sigma_c \neq
0$ bounded by the lines $2-K+\sigma K^2=0$ and $\sigma=1/8$.
Furthermore, the line $\sigma=2/K^2$ manifests a phase transition
corresponding to a change in the relation $\sigma_c \sim \epsilon
^{2/z}$ where
\begin{eqnarray}
z=2-K+\sigma K^2&  \mbox{\hspace{15mm}}& \sigma<2/K^2 \nonumber\\
z=K(\sqrt{8\sigma}-1)&     & \sigma>2/K^2 \, . \label{z}
\end{eqnarray}
This transition occurs as rare regions of the sample rather than
typical ones start dominating the behavior \cite{carpentierxy}, as
can be also seen from (\ref{sigmac}). For $\sigma<2/K^2$ one can
discard denominators (as well as the $\epsilon^2$ term) which
immediately yields (\ref{z}). For $\sigma>2/K^2$ the average over
the random fugacity $\omega$ is dominated by the tail of the distribution
and the right hand side of (\ref{sigmac}) can be approximated by
\cite{integral}:
\begin{eqnarray}
\text{Proba}(\omega + u + \ln \epsilon > 0) \approx \exp(\frac{(K
+ 2 \frac{\ln \epsilon}{\ln \sigma_c })^2}{8 \sigma K^2} \ln
\sigma_c)
\end{eqnarray}
yielding the strong disorder form of (\ref{z}).

The above results now allow to compute straightforwardly the {\it
disorder averaged} DOS. Indeed we can identify the disorder
average of Eq. (\ref {rho}) as
 $\overline{\rho_{\epsilon}(0)}=\frac{\partial}{\partial\epsilon}
F_{var}/\pi L^2$ with
the overline denoting disorder average. This yields the replica average
\begin{equation}
\overline{ \rho_{\epsilon}(0)}=\frac{1}{\pi}\frac{\partial}{\partial\epsilon}
\langle\sum_{{\bf n}}Y[{\bf n}]\exp (i{\bf n} \cdot
{\mbox{\boldmath $\theta$}})\rangle \,. \label{rhobar}
\end{equation}
Using Eq. (\ref {sigmac}) for $\sigma_c$ and a corresponding equation
for $\sigma_0$ we find $\overline{ \rho_{\epsilon}(0)}\sim
(\sigma_c+\sigma_0)/\epsilon \sim \epsilon^{2/z-1}$. Note that Eq. (\ref{rhobar})
has beyond the $\langle\cos\theta\rangle$ term all the higher order
terms in $\epsilon$ as generated by RG. The result differs from just
the $\langle\cos\theta\rangle$ average in the strong disorder regime,
$\sigma>2/K^2$.

In this derivation we have used that $L$ is large compared with
the correlation length $1/\sqrt{\sigma_c}\sim\epsilon^{-1/z}$ so
that integration cutoffs are determined by $\sigma_c$. Hence our
result is that
\begin{equation}
\overline{ \rho_{\epsilon}(0)}\sim \epsilon^{2/z-1}
\mbox{\hspace{15mm}} \epsilon>L^{-z} \label{rho1}
\end{equation}
exhibits a phase transition at $\sigma=2/K^2$. From
(\ref{integral}) this also implies that the DOS also exhibits a
transition, with $\overline{ \rho(E)}\sim E^{2/z-1}$. The
condition $\epsilon>L^{-z}$ can be interpreted as the typical
level spacing $\Delta E=1/\overline{ \rho_{\epsilon}(0)}L^2$ being
small, $\Delta E<\epsilon$, so that levels overlap and the DOS is
smooth at finite $L$.

Our conclusion for the free random Dirac Hamiltonian
(\ref{hdirac}), obtained by setting $K=1$, is that there is a
phase transition at $\sigma =2$. We find:
\begin{eqnarray}
& \rho(E) \sim E^{(1-\sigma)/(1+\sigma)} & \quad \text{for}\,\, \sigma
<2 \\
& \rho(E) \sim E^{(3-\sqrt{8\sigma})/(\sqrt{8\sigma}-1)}& \quad
\text{for}\,\, \sigma >2
\end{eqnarray}
 The value obtained for the
threshold coincide with the one for the exact zero energy
eigenstate \cite{chamon,castillo}. Thus at this value of disorder
a sharp change of behaviour also occurs in the DOS. A more
detailed treatment reveals that in the strong disorder phase there
are logarithmic prefactors to the DOS, as was also the case in
\cite{freezing}. For $\sigma > 2$ one has:
\begin{eqnarray}
&& \rho(E) \sim E^{2/z-1} |\ln E|^{-\psi}
\end{eqnarray}
where we find for $E=i \epsilon$ that $\psi=\frac{1}{2} \gamma
\frac{\sqrt{8\sigma}}{\sqrt{8\sigma}-1}$, with from
\cite{integral} $\gamma=1/2$.

As mentioned above, using the RG of \cite{carpentierxy} yields the
same result (\ref{z}) for the dynamical exponent $z$. In fine,
this is a result about the true scaling dimension $z$ of the
$\epsilon \cos(\theta)$ operator (noted $\Delta_{typ}$ in the
conclusion of \cite{carpentierxy}) being different, in the strong
disorder regime, from the naive one (noted $\Delta$ there), as
occurs in the XY model (see discussion there). The RG treatment is
expected to change the value of the exponent $\psi$ of the
logarithmic corrections (i.e. $\gamma$ is expected to change to
$\gamma=3/2$ \cite{carpentierpart}).

\begin{figure}[thb]
\centerline{\fig{8cm}{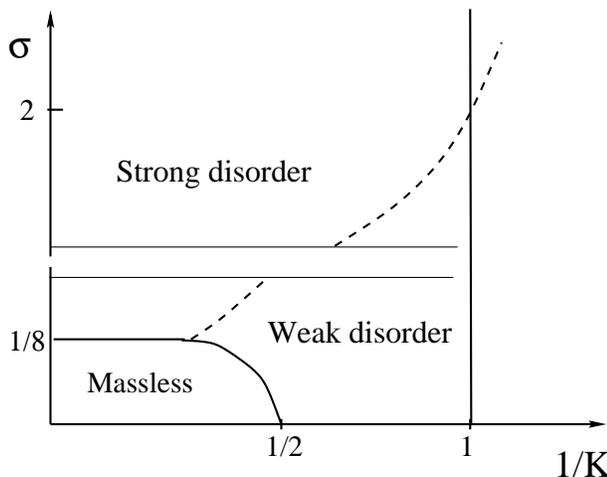}}
\caption{ \narrowtext
Schematic phase diagram for the generalized Dirac problem (e.g. single layer
disordered Coulomb gas). $\sigma$ is the strength of the random gauge disorder.
$K$ is an interaction parameter so that $K=1$ corresponds to free fermions.
The full line is the phase boundary above which single charges become relevant
and below which the model is massless.
The dashed line indicate the freezing transition between weak and strong disorder
regimes. \label{fig1}}
\end{figure}

\subsection{Finite size regime}

Let us now characterize some aspects of the fluctuations of the
DOS in a finite size system. Within the variational method described
above one sees that for $\epsilon>L^{-z}$ the system is too small to generate the mass
$\sigma_c$ hence $\rho_{\epsilon}(0)\sim\langle\cos\theta\rangle_0$
with $\sigma_c=\sigma_0=0$ in Eq. (\ref{H0}), i.e.
\begin{equation}
\overline{ \rho_{\epsilon}(0)}\sim L^{-K+\sigma K^2} \mbox{\hspace{15mm}}
\epsilon<L^{-z}  \label{rho2} \,.
\end{equation}
Since the system is effectively massless we expect
significant fluctuations. In the following we consider a different
approach for the  $\epsilon<L^{-z}$ case which will clarify the nature
of disorder average.

We proceed to evaluate the DOS by a direct expansion in $\epsilon$. At $\epsilon = 0$ a direct
evaluation of the Gaussian average over $\theta$ in a given sample (assuming periodic boundary
conditions for the resulting potential $V({\bf r})$) yields \cite{footnoteboundary}:
\begin{equation}
\langle cos\theta({\bf r})\rangle =e^{-K\ln L}[e^{- U({\bf
r})}+e^{U({\bf r})}]/2
\label{u}
\end{equation}
where, in Fourier space, $U(q)=2 K V(q) = (2 K/q^2)(iq_x
A_y-iq_yA_x)$ with correlation of the form
\begin{equation}
\overline{(U({\bf r}) - U(0))^2} = 4 \sigma K^2 \ln r \label{corr}
\end{equation}
The density of states thus takes the form
\begin{equation}
\rho_{\epsilon=0}(0) = \frac{1}{L^2} \sum_{\bf r} e^{-K \ln L}[e^{-U({\bf
r})}+e^{U({\bf r})}]/2 = \frac{1}{L^2} Z_L
\label{relation}
\end{equation}
of the partition function $Z_L$ of a single $\pm$ vortex in a
logarithmically correlated random potential, known to be
related to the one of a directed polymer on a
Cayley tree \cite{tang_xy_lowtemp,carpentierpart}. A simple average of the partition
function $Z_L$, i.e. ${\bar \rho_{\epsilon}(0)}$
indeed leads to Eq. (\ref{rho2}), however, as is well known in the directed
polymer problem only the logarithm of the partition function $\ln Z_L$ is self averaging.
This immediately yields \cite{cook,derrida,carpentierpart} our result for the typical DOS at finite size:
\begin{equation}
\rho_{typ}(0) \sim L^{-2+z}
\end{equation}
with $z$ given by Eq. (\ref{z}).

To identify the role of $\epsilon$ we consider
the first order terms $\epsilon \int d^2r'\langle \exp [\pm \theta
({\bf r})\pm \theta({\bf r}')\rangle$. The typical value of
each of these terms scales as the typical value of
$Z_L^2/L^2$ (for opposite charges it is true in the massive
phase we are interested in).
This allows to identify a crossover function $f(x)$, where
\begin{equation}
\rho_{typ}(0)\sim L^{-2+z}+\epsilon L^{-2+2z}+... =L^{-2+z}f(\epsilon
L^z)
\end{equation}
with $f(x)=1+x$ at $x\rightarrow 0$. For $x\gg 1$ we can recover
Eq. (\ref{rho1}) if the crossover function satisfies
 $f(x)\sim (1/x)^{1-2/z}$, i.e. the typical value $\rho_{typ}(0)$
 crosses over to the average ${\bar \rho_{\epsilon}(0)}$ at $\epsilon
>L^{-z}$, with both limits exhibiting a phase transition.

This statistics can be described in a simple phenomenological picture. A finite $\epsilon$
provides a length scale (the vortex separation) and in effect cuts the system in independent
pieces of sizes $L_\epsilon=\epsilon^{-z}$. One has thus roughly:
\begin{equation}
\rho_\epsilon(0) = \frac{1}{L^2} \sum_{i=1}^{(L/L_\epsilon)} Z^{(i)}_{L_\epsilon}
\end{equation}
where the random variables $Z^{(i)}_{L_\epsilon}$ are independent with
a lognormal distribution. For large $(L/L_\epsilon)$, however, the above sum acquires a
normal distribution.  A similar picture was used to describe the
related random diffusion problem, where the local first passage times
are analogous to the local DOS in the present problem, and an external force
produces a finite length scale. Analysis of the various regimes is described
there and are expected to be quite similar here.

\subsection{relation to random diffusion models}

It is instructive to compare our results to the one
obtained for random diffusion problems.
As mentionned in the introduction, general random Dirac problems
can be mapped onto random diffusion operators, which in general
may involve two species, with absorption, creation and transformation.
It is particularly simple in the case $W=0$ (random vector potential
alone) where it maps onto a random Fokker Planck diffusion operator
of the type:
\begin{eqnarray}
H_{FP} P = \nabla^2 P - \nabla (F_T + F_L) P = - E' P
\end{eqnarray}
which describes the Langevin diffusion of a particle, $\partial_t P = H_{FP} P$
where $P({\bf r})$ is the probability that the particle is at point ${\bf r}$
at time $t$, in the presence of both a potential random force $F_L=- \nabla U$
and a transverse one (a random drift), with $div F_T=0$. Equivalently, setting
$P = e^{- U/2} \psi$:
\begin{eqnarray}
H'_{FP} \psi = (\nabla^2  - F_T \nabla - (\nabla V)^2 + \nabla^2 V) \psi  = - E' \psi
\end{eqnarray}
with $V=U/2$ ($K=1$ here). The operators $H_{FP}$ and $H_{FP}'$ have the same
spectrum. In two dimension, taking the square of (\ref{hdirac}) with $W=0$ yields:
\begin{eqnarray}
&& - H_D^2 = \nabla^2 - (A_x^2 + A_y^2) + \sigma_z (\partial_y A_x - \partial_x A_y) \\
&& - 2 \lambda i A \cdot \nabla - i \nabla \cdot A
\end{eqnarray}
with $\lambda=1$ identical to $H'_{FP}$ with $A_x=\partial_y V$ and $A_y=- \partial_x V$ (and $V \to -V$
for the other component of the spinor) and $F_T = 2 i A$. This is thus Arrhenius diffusion
in the random potential $U$ with an additional {\it imaginary} random drift \cite{old}. The
diffusion dynamical exponent $z_d$ associated with $H'_{FP}$ should thus be simply
$z_d = 2 z$. Note that all the operators obtained by varying $\lambda$
have identical ground state wavefunction, $\psi \sim e^{-V}$
since the additional drift term does vanish in the ground state
(in the diffusion context it means that the drift is along equipotentials of $U$).
It is thus reasonable to expect that each of these models are described by a
line of fixed points and that they all do exhibit a freezing transition
for any value of $\lambda$ {\it at the same value} of $\sigma=\sigma_{th}=2$.

In the absence of this additional drift (i.e. setting $\lambda=0$), the problem reduces to the one studied in
\cite{freezing} where indeed it was found that there is also a freezing transition in the
dynamical exponent $z_d$ in $d=1$ and $d=2$ with (assuming conventional dynamical scaling):
\begin{eqnarray}
&& z_d(\lambda=0) = 2 + 2 (\sigma/\sigma_{th}) \qquad \sigma < \sigma_{th} \\
&& z_d(\lambda=0) = 4 \sqrt{\sigma/\sigma_{th}}\qquad \sigma > \sigma_{th}
\end{eqnarray}
Although it does indeed exhibit a freezing transition at the same threshold $\sigma_{th}=d$
one sees from () that $z_d(\lambda=0) \leq z_d(\lambda=1)=2 z$, i.e. the imaginary drift slow down
the diffusion, presumably through interference effects. It would be of interest to
determine $z_d(\lambda)$ as well as to study freezing transitions in a generalized class
of these diffusion models in two dimensions.

It is possible to consider various one dimensional restriction of the Dirac model, e.g.
the so called supersymmetric quantum mechanics which also exhibits
band center delocalization \cite{comtet98,balentsfisher}. With a log correlated $U(x)$ this model was studied analytically
in \cite{freezing}, thus we know in that case the exact $z=z_d/2$ dynamical exponent of the
random Dirac operator.

\section{Full quantum Hall problem}

\subsection{one layer problem: scaling of the DOS}

Finally, we consider the Dirac model where the scalar random
potential in Eq. (\ref{SD}) is retained, which describes the full
quantum Hall system. We will first determine the DOS first at zero
energy, and later around zero energy.

In presence of a random scalar potential one has $Y[{\bf n}]\sim
(iW-\epsilon)^{\sum n_a^2}$ in Eqs. (\ref{H},\ref{rhobar}). We can
safely set $\epsilon=0$ in the definition $\rho_{\epsilon}(0)\sim
\partial F_{var}/\partial \epsilon$ since $W({\bf r})$ provides a mass parameter. The variational
method is similar to the previous case, except that $\epsilon$ is
replaced by $iW({\bf r})$ in Eq. (\ref{sigmac}). Since the
integral is dominated by large $u$ and $\omega$ (for $\omega >0$),
 it has the form:
\begin{eqnarray}
&& \sigma_c=\langle
\frac{iWe^{u+\omega}}{(1+iWe^{u+\omega})^2}\rangle_{\omega,W}
\\
&& =
\langle \frac
{2W^2e^{2u+2\omega}}{(1+W^2e^{2u+2\omega})^2}\rangle_{\omega,W}
\end{eqnarray}
using the $\pm$ symmetry of the $W$ average \cite{integral}. The
latter form is equivalent to the previous integral Eq. (\ref{z})
with $\epsilon$ replaced by the disorder average $\langle
W^2\rangle=\delta$ (for the starting QH system $\delta\sim\sigma$)
 and $K$ is replaced by $2K$. Hence $\sigma_c\approx\delta^{2/z'}$ where now
\begin{eqnarray}
z'= 2-2K+4\sigma K^2&  \mbox{\hspace{15mm}}& \sigma<1/2K^2
\nonumber\\
 z'= 2K(\sqrt{8\sigma}-1)& &
\sigma > 1/2K^2 \, . \label{z'}
\end{eqnarray}
The DOS at zero energy can be written as
\begin{equation}
\overline{\rho(E=0)} \sim
\langle\cos\theta\rangle\sim\langle\frac{e^{u+\omega}}
{1+W^2e^{2u+2\omega}}\rangle_{\omega,W}\sim\delta^{\alpha}
\end{equation}
where
\begin{eqnarray}
\alpha=\frac{2}{z'} - \frac{z}{z'}  \label{alpha}
\end{eqnarray}
Since $z'$ has a transition at $\sigma=1/2K^2$ the DOS has {\it
two transitions}, at $\sigma=1/2$ and at $\sigma=2$ (for $K=1$).
The exponent $\alpha$ in Eq. (\ref {alpha}) is the one expected
from a scaling form $\rho_{\epsilon,\delta}(0)= \delta^{\alpha}
g(\epsilon/\delta^{z/z'})$ which connects with the $\delta=0$ case
solved in Section III (which requires $g(x)\rightarrow 1$ at
$x\rightarrow 0$ and $g(x)\sim x^{\alpha z'/z}$ at $x\rightarrow
\infty$).
 As we will see
below there are however three phases, each with a different
scaling function $g(x)$. Note that $z'/z$ increases from $0$ at
small $\sigma$ to $z'/z=2$ at $\sigma=2$ and remains equal to this
value for stronger disorder.

We have thus shown that the DOS is finite at the quantum Hall transition, which
is a well established result \cite{evangelou}. In addition however, we have determined
the scaling of the zero energy density of states with the scalar
potential strength in the random Dirac problem. Note that in the
previous case with $\epsilon \rightarrow 0$ renormalization of $K$
and $\sigma$ was of higher order in $\epsilon$ and could be
neglected. Here, however, a finite $\langle W^2({\bf r})\rangle$
renormalizes both $K$ and $\sigma$, with $\sigma$ flowing to
stronger values. This does not spoil our result for the exponent
$z'$ in the limit $\delta \to 0$ however, as it would also yield
only subleading corrections in $\delta$.

The physics of the problem becomes more apparent when one
notes that the random scalar potential produces (see e.g. Eq. (\ref{SB}) )
a random (imaginary) fugacity with a random sign. One can then
again either extend the RG of \cite{carpentierxy} to this situation
or consider the extreme dilute limit (single vortex) as in Eq. (\ref{u}).
Both considerations lead immediately to a mapping onto the directed
polymer on the Cayley tree with random Boltzmann weights {\it of random
sign } (in the bare model the random sign acts only at the leafs of the tree
but since both signs have equal probability one easily sees that this is
equivalent to a random sign on each branch of the tree). This model was
also solved in \cite{derridacomplex} and is known to
 indeed exhibit a transition
at half the value of the same sign problem, due to interference
effects which in effect bind two replicas. The value of the $z'$
exponent obtained by this method is identical to the one given
above (Eq. (\ref{z'})) . It is remarkable, and encouraging, that
the variational method also captures this physics.

We can now extend these considerations and obtain, in implicit
form, the crossover function which describes the DOS $\rho(E)$ in
the small $\delta$ limit. We will obtain explicitly
$\overline{\rho_\epsilon(0)}$, from which $\overline{\rho(E)}$ can
be extracted inverting (\ref{integral}).

The equation which determines $\sigma_c$ and the DOS is now:

\begin{eqnarray}
&& \sigma_c = \langle\int d\omega\,
\frac{\epsilon e^{u + \omega} + (W^2 + \epsilon^2) e^{2 u + 2 \omega}}{
1 + 2 \epsilon e^{u + \omega} + (W^2 + \epsilon^2) e^{2 u + 2 \omega}} \rangle_{\omega,W} \\
&& \overline{\rho_\epsilon(0)} =  \langle\int d\omega\,
\frac{e^{u + \omega}+\epsilon e^{2u+2\omega}}{1 + 2 \epsilon e^{u + \omega} + (W^2 + \epsilon^2) e^{2 u + 2 \omega}}
\rangle_{\omega,W}
\end{eqnarray}
the first line is really $\sigma_c + \sigma_0$ but we can use that
$\sigma_0$ is subdominant and only dominant exponents for the
$\omega >0$ are retained \cite{integral}. We immediately see that
there are several regimes, according to whether one can neglect
all denominators (weak disorder), or whether the averages will be
dominated by the rare events where either the terms proportional
to $W^2$ or to $\epsilon$ or both, are of order one. There are in
fact three phases, in each of them scaling holds with different
scaling functions $f$,$g$,${\cal R}$:
\begin{eqnarray}
&& \sigma_c  = \delta^{2/z'} f(\epsilon/\delta^{z/z'})  \label{f} \\
&& \overline{\rho_\epsilon(0)} = \delta^{(2-z)/z'}
g(\epsilon/\delta^{z/z'})
\end{eqnarray}
from which one can extract the DOS scaling function
\begin{eqnarray}
&& \rho(E) = E^{2/z - 1} {\cal R}(E/\delta^{z/z'})
\end{eqnarray}
determined implicitly by the relation:
\begin{eqnarray}
&& g(x) = \frac{1}{\pi} \int dy y^{2/z - 1} {\cal R}(y)
\frac{x}{x^2 + y^2}
\end{eqnarray}

{\it weak disorder phase } $\sigma < 1/2K^2$:

Neglecting all denominators the equation for $\sigma_c$ and
$\overline{\rho_\epsilon(0)}$ become:
\begin{eqnarray}
&&  \overline{\rho_\epsilon(0)} =  \sigma_c^{1 - z/2} \\
&& 1 = \epsilon \sigma_c^{- z/2} + \delta \sigma_c^{- z'/2}
\end{eqnarray}
which yield the scaling functions in implicit form:
\begin{eqnarray}
&&  x = f(x)^{z/2} (1 - f(x)^{-z'/2})  \\
&&  g(x) = f(x)^{1 - z/2}
\end{eqnarray}
possibly the exact ones, up to prefactors.

{\it strong disorder phase I } $1/2K^2 < \sigma < 2 K^2$:

There one can still neglect denominators in averages involving the
terms proportional to $\epsilon$ but not in the one involving the
terms proportional to $W^2$. One finds:
\begin{eqnarray}
&&  \overline{\rho_\epsilon(0)} \sim  \sigma_c^{1 - z/2} \\
&& \sigma_c = \epsilon \sigma_c^{1 - z/2} + \sigma_c^{ (K +
\frac{\ln \delta}{\ln \sigma_c})^2/(8 \sigma K^2)}
\end{eqnarray}
Simple expansion shows that scaling still holds but the scaling
functions are now implicitly given by:
\begin{eqnarray}
&&  x = x = f(x)^{z/2} (1 - f(x)^{-\tilde{z}'/2}) \\
&&  \tilde{z}' = 4(1 - \frac{1}{\sqrt{8 \sigma}}) \\
&&  g(x) = f(x)^{1 - z/2}
\end{eqnarray}
This scaling function is accurate only up to logarithmic
prefactors. A more accurate form is $ \sigma_c \sim \delta^{2/z'}
| \ln \delta |^{- 2 \gamma/\tilde{z}'} f(\epsilon \delta^{-z/z'} |
\ln \delta |^{2 \gamma/\tilde{z}'} )$ with$\gamma=1/2$ but
$\gamma$ is likely to be corrected upon a more careful RG
treatment.

{\it strong disorder phase II } $2/K^2 < \sigma $:

At even stronger disorder, since $z'=2 z$, we expect the scaling
region to be $\epsilon^2 \sim \delta$. To show that this is the
case and to get some approximation for $f(x)$ we notice that the
equation for $\sigma_c$ can be approximated in the scaling region
by:
\begin{eqnarray}
&& \sigma_c \approx \text{Proba}(\epsilon e^{u+\omega} +
(\epsilon^2 + \delta) e^{2 (u+\omega)}  > 1 )
\end{eqnarray}
Solving the quadratic equation in $e^{u+\omega}$ yields that
$\sigma_c$ is indeed of the form (\ref{f}), with some form for
$f(x)$, which here is approximate. We have not attempted to solve
more precisely for $f(x)$ or $g(x)$ in this phase.

Finally note that this crossover can also be studied at finite size, and is there
complicated as it will probably be described as in \cite{derridacomplex}
by a non trivial phase diagram.

\subsection{Transport in the quantum Hall system : the two layer problem}

We address now the more difficult problem of describing transport
and localization in the QH system. Transport is derived by a
disorder average of advanced and retarded propagators, hence two
partition sums corresponding to the action of Eq. (\ref{SB}) with $\pm
\epsilon$. The problem is then of two layers with fields
$\theta_{\uparrow}$ and $\theta_{\downarrow}$ and common disorder
${\bf A}({\bf r})$ and $W({\bf r})$. The role of $\pm \epsilon$ is
to determine the proper ground state near which the variational
method applies, i.e. for $-\epsilon$ we shift
$\theta_{\downarrow}\rightarrow \theta_{\downarrow}+\pi$ so that
the nonlinear terms become
\begin{equation}
\exp\{-\sum_a[(-iW+\epsilon)\cos\theta_{\uparrow,a}+
(iW+\epsilon)\cos\theta_{\downarrow,a}]\}
\end{equation}
where $a=1,...,m$ are replica indices for each layer and we have
redefined here $W/\pi\alpha,\,\epsilon/\pi\alpha\rightarrow W,\,
\epsilon$, respectively. Expansion in powers of $iW\pm\epsilon$
and identifying the dominant terms of the form $\exp (i{\bf n}
\cdot {\mbox{\boldmath $\theta$}})$ yields
\begin{equation}
{\cal H}_{int}= \sum_{{\bf n}}(-iW)^{\sum_an^2_{\uparrow a}}
(iW)^{\sum_an^2_{\downarrow a}} \exp (i{\bf n} \cdot
{\mbox{\boldmath $\theta$}}) \label{Hint}
\end{equation}
with $\epsilon$ now set to zero, ${\bf n}$ is now a vector of $2m$
entries $(n_{\uparrow 1},...n_{\uparrow m},n_{\downarrow
1},...,n_{\downarrow m})$ and similarly for ${\mbox{\boldmath
$\theta$}}$.

The Gaussian part of the Hamiltonian can be written as
\begin{eqnarray}
&& {\cal H}'=\sum_{{\bf q},a} \frac{q^2}{8\pi K}(|\theta_{+,a}({\bf
q})|^2+ |\theta_{-,a}({\bf q})|^2) \nonumber \\
&& +\sum_{{\bf q},a,b}\frac{\sigma
q^2}{4\pi} \theta_{+,a}({\bf q})\theta_{+,b}(-{\bf q})
\end{eqnarray}
where $\theta_{\pm a}=(\theta_{\uparrow a}\pm \theta_{\downarrow
a})/\sqrt{2}$. Since only the $\theta_{+}$ mode is affected by the
common disorder ${\bf A}({\bf r})$ we expect $\theta_{\pm}$ to
have distinct self masses. The variational Hamiltonian is then
\begin{equation}
{\cal H}_0={\cal H}'+\frac{1}{2}\sum_{{\bf q}
,\pm,a,b}[\sigma_c^{\pm} |\theta_{\pm a}({\bf
q})|^2\delta_{ab} +\sigma_0^{\pm}\theta_{\pm a}({\bf q})\theta_{\pm
b}(-{\bf q})] \,.
\end{equation}
The propagators of the $\pm$ modes are used to define
\begin{eqnarray}
&& \sum_{{\bf q}}\langle\theta_{\pm a}({\bf
q})\theta_{\pm
b}(-{\bf q})\rangle_0 = -4u_{\pm}\delta_{ab}-2A_{\pm}\nonumber\\
u_{\pm}&=&-(K/4)\ln (\Delta_c/4\pi K\sigma_c^{\pm})\nonumber\\
A_+&=& \sigma K^2\ln(\Delta_c/4\pi
K\sigma_c^+)+K\sigma_0^+/2\sigma_c^+ - \sigma K^2\nonumber\\
A_- &=& K\sigma_0^-/2\sigma_c^-  \,.
\end{eqnarray}
We write the interaction Eq. (\ref {Hint}) in the form
\begin{eqnarray}
&&\langle{\cal H}_{int}\rangle_0=\langle (-iW)^{\sum_an^2_{\uparrow
a}}(iW)^{\sum_an^2_{\downarrow a}}\nonumber \\
&&\exp[u_+\sum_a(n_{\uparrow a}+n_{\downarrow a})^2+
\omega_+\sum_a(n_{\uparrow a}+n_{\downarrow
a}) \nonumber \\
&& + u_-\sum_a(n_{\uparrow a}-n_{\downarrow a})^2 +
\omega_-\sum_a(n_{\uparrow a}-n_{\downarrow a})]\rangle_{\omega}
\label{Hav}
\end{eqnarray}
where the $\omega$ average reproduces the required form with
$A_{\pm}(\sum_a n_{\uparrow a}\pm n_{\downarrow a})^2$,
\begin{equation}
\langle ...\rangle_{\omega} =\int\int \exp
[-\frac{\omega_+^2}{2A_+}-\frac{\omega_-^2}{2A_-}]
\frac{d\omega_+d\omega_-}{2\pi\sqrt{A_+A_-}}  \,.
\end{equation}
The sum in Eq. (\ref{Hav}) can be written as
 $\langle{\cal H}_{int}\rangle_0=\langle Z^m\rangle_{\omega}$
with
\begin{eqnarray}
&& Z= 1-iWe^{u_++u_-+\omega_++\omega_-}+iWe^{u_++u_-+\omega_+-\omega_-}
\nonumber \\
&& -iWe^{u_++u_--\omega_+-\omega_-}+iWe^{u_++u_--\omega_++\omega_-}\nonumber\\
&& + W^2e^{4u_++2\omega_+}+W^2e^{4u_-+2\omega_-} \nonumber \\
&& +W^2e^{4u_--2\omega_-}
+W^2e^{4u_+-2\omega_+}
\end{eqnarray}
The masses are to be found by minimizing the variational free
energy $F_{var}=F_0 + \langle H-H_0 \rangle_0$, as in section III A.
We expect to find a massless solution, e.g. $\sigma_c^-=0$. This
is indeed a possible solution with $u_-\rightarrow -\infty$ and
\begin{equation}
Z=1+W^2e^{4u_++2\omega_+}+W^2e^{4u_+-2\omega_+} \,.
\end{equation}
$W^2$ can be replaced by its average and then this has the same
structure as the single layer problem of section IIIA with the
replacement $K\rightarrow 2K$, i.e. the phase diagram is the same
as Fig. 1 with the $1/K$ axis replaced by $1/2K$. The starting line
$K=1$ is now tangent to the phase boundary at $\sigma=0$. However, for
$\sigma > 0$ $\theta_+$ is massive.

To find the QH localization exponent we introduce the mass term which
corresponds to
\begin{equation}
\Delta_d(\sin\theta_{\uparrow}-\sin\theta_{\downarrow})=
2\Delta_d\cos(\theta_+/\sqrt{2})\sin(\theta_-/\sqrt{2})
\end{equation}
Note the opposite signs due to the shift of $\theta_{\downarrow}$
 as required by
the sign of $\epsilon$. Since $\theta_+$ is massive, $\Delta_d$ can be
replaced by ${\tilde\Delta}_d=2\Delta_d\langle\cos(\theta_+/\sqrt{2})\rangle$
and defining $\theta=\theta_-/\sqrt{2}+\pi/2$ leads to an effective
Hamiltonian with
\begin{equation}
H_{eff}=\int d^2 r [\frac{1}{4\pi K}|\nabla\cdot\theta({\bf r})|^2 -
\tilde{\Delta}_d\cos\theta({\bf r}) ]
\end{equation}
This system has a correlation length $\xi$, related to a mass
 $\xi^{-2}\sim \Delta_d^{2\nu}$ which from first order RG is
 $\nu=2/(4-K)$. For the starting QH
problem with $K=1$ the localization exponent is then $\nu=2/3$
(the numerically known value is $\approx 2.3$). We expect,
however, that $K$ is RG driven to a different value. This is
beyond the variational scheme which gives reliable exponents only
for small ($W, \epsilon$) couplings (but arbitrary $\sigma,K$), as
in the $\epsilon \rightarrow 0$ case. E.g., to 2nd order in
replicated sine Gordon RG, the most relevant operator $W\cos
(\theta_a+\theta_b)$ yields
\begin{eqnarray}
\frac{dW}{d\ell}&=&(2-2K+4\sigma K^2)W\nonumber\\
\frac{dK^{-1}}{d\ell}&=&-2(K-2\sigma K^2)W^2\nonumber\\
\frac{d\sigma}{d\ell}&=&(K-2\sigma K^2)W^2
\end{eqnarray}
which shows that indeed $K$ increases at weak disorder.

We have also looked for other solutions of the variational scheme
of the form $\sigma_c^-\sim(\sigma_c^+)^{\alpha}$ and found that
only $\alpha=1$ exists. This corresponds to decoupled layers with
the phase diagram of Fig. 1 ($K$ replaced by $2K$). Hence at $K=1$
this is a massive phase and does not correspond to the QH problem.

Our method shows that a subset of the degrees of freedom can form a massless
phase and we hope that it will stimulate further progress.

\section{conclusion}

In conclusion, we have derived an exact formulation of two
dimensional random Dirac fermions in terms of a random sine Gordon
model, or equivalently a disordered Coulomb gas. The DOS of the
Dirac fermion system identifies with the expectation value of a
cosine operator in the sine Gordon model ( equivalently the charge
fugacity in the Coulomb gas ) and the dynamical exponent $z$ as
its scaling dimension. We found that at zero energy and with
random vector potential only the Dirac system maps onto the random
gauge XY model with infinitesimal vortex fugacity (dilute limit of
the CG). Using methods and results from previous studies of the XY
model, we have computed the exact dynamical exponent $z$ for the
random vector potential model at any disorder, and thus obtained
the critical behavior of the DOS around zero energy. We found that
it exhibits a transition at the same threshold value than the
previously known transition in an exact ground state wave
function. It shows that all eigenstates near the band center are
affected by this transition. The physics of this transition is
closely related to the freezing transition in the XY model in the
limit where the vortex core energy is taken very large. As we show
here the density of states in a finite size sample becomes broadly
distributed with typical values scaling differently than average
ones with the system size. This corresponds to the eigenstates
being peaked around some few centers in the sample.

It is likely that similar freezing phenomena are of importance in
a broader class of two dimensional disordered models. They were
recently found to occur in random diffusion models, for instance
in the problem of random Arrhenius diffusion of a particle in a
logarithmically correlated potential. We have analyzed the
similarities and differences of the transitions which occur in the
dynamical exponents $z$ of both models. As was shown in
\cite{freezing} strong disorder renormalization group captures the
dynamical behaviour in the glass phase, which suggests that it
could be used to study the present problem as well.

It is of utmost importance to understand what are the consequences
of this transition when a random potential is added, corresponding
to the quantum Hall system. As a first step in that direction we
have determined the scaling of the DOS (which is then finite at
zero energy) when the additional random potential is small. There
we found two transitions, one at the same value than the random
vector potential model, the other at a much smaller value of
disorder. It would be nice to know whether these transition lines
extend away from the random vector potential fixed line. Numerical
checks of our predictions as well as a numerical calculation of a
glass order parameter (e.g. $\sum_x |\psi(x)|^4$ ) in the full
model would help understand these issues.

We believe that such freezing phenomena, originally studied in the
context of disordered Coulomb gas and XY models, will also affect
a broader class of disordered fermion models in two dimensions.
This can be studied systematically by extending the bosonization
approach introduced in the present paper, and for instance,
searching for all perturbations around the random gauge fixed
plane and computing their (non trivial) scaling dimension (here it
was done only for the vortex fugacity operator). In particular it
is of interest to know whether the non linear sigma models studied
in \cite{nlsigma} also exhibit freezing phenomena. They are indeed
generalizations of the Liouville model which captures the single
vortex problem and does exhibit a freezing transition. Results are
already known in related cases. For instance it was shown in
\cite{giam} that if one adds pinning disorder to the random gauge
XY model the vortex density (DOS) acquires a $\sim exp( - |\ln
\epsilon|^{2/3} )$ dependence in the vortex core energy $\ln
(1/\epsilon)$ with a non trivial exponent.

Finally, we have formulated the problem of the transport in the
quantum Hall system as two coupled random sine Gordon models. We
have applied the variational method, which should capture some of
the non perturbative effects. We have analyzed the system using in
the plane of two (dimensionless) parameters and were able to find
a massless phase. Though qualitatively encouraging, to obtain the
quantitative characteristics of this phase requires to incorporate
in a more precise way additional renormalizations of these two
parameters. It is tantalizing that possible values of these
renormalized parameters (consistent with numerics) seem to lie
within the region near the glass phase boundary.

We thank Bernard Derrida and Leon Balents for useful discussions. B.H. acknowledges support
by THE ISRAEL SCIENCE FOUNDATION founded by the Israel Academy of Science and Humanities.

\unecol


\begin{references}
\bibitem{footnoteboundary}
Let us note that (\ref{relation}) makes sense only because
we have not detailed the correspondence at finite size between the
fermion and boson path integral (had we done that it should have
given an infinite result e.g. if the boundary conditions are consistent
with the exact eigenstate). Thus it should be already, in a given sample,
considered as an average over some boundary conditions.
\bibitem{integral}
In the strong disorder regime the exact behavior of these
integrals for large logarithms ($\sigma_c$ small) are easily
obtained through a shift of variable \cite{carpentierxy}. For
instance the right hand side of (\ref{sigmac}) is $\approx
\frac{I}{\sqrt{2 \pi A} } \exp{(- (u + \ln \epsilon)^2/2A})$. $I$
is a constant of order one, given by the integral $I= 2 \int
d\omega' e^{- \mu \omega'} \frac{e^{\omega'}}{(1 +
e^{\omega'})^2}$ with $\mu=-(u + \ln \epsilon)/A \to ( K/2 +
\frac{\ln \epsilon}{\ln \sigma_c })/2 \sigma K^2$ which is
convergent for $\mu<1$, i.e. here for $\sigma > 2 K^2$. The factor
of $2$ arises from the two regions, symmetric in $\omega \to -
\omega$, which contribute to the integral. Since we do not keep
track here of (non universal) prefactors (which can be absorbed in
change in cutoff) we can safely drop the terms with $e^{-\omega}$
in the following.
\end{references}
\end{document}